\documentclass[preprint,showpacs,aps]{revtex4}
\usepackage{graphicx}% Include figure files
\begin{document}
\def\be{\begin{equation}}
\def\ee{\end{equation}}
\def\bea{\begin{eqnarray}}
\def\eea{\end{eqnarray}}
\title{Geometric description of BTZ black holes thermodynamics}
\author{Hernando Quevedo}
\email{quevedo@nucleares.unam.mx}
\affiliation{ Instituto de Ciencias Nucleares\\
Universidad Nacional Aut\'onoma de M\'exico  \\
AP 70543, M\'exico, DF 04510, MEXICO}
\author{Alberto S\'anchez}
\email{quevedo@nucleares.unam.mx}
\affiliation{ Instituto de Ciencias Nucleares\\
Universidad Nacional Aut\'onoma de M\'exico  \\
AP 70543, M\'exico, DF 04510, MEXICO}
\begin{abstract}
% Text of abstract

We study the properties of the space of thermodynamic 
equilibrium states of the Ba\~nados-Teitelboim-Zanelli (BTZ)
black hole in (2+1)-gravity. We use the formalism 
of geometrothermodynamics to introduce in the space of
equilibrium states  a $2-$dimensional 
thermodynamic metric whose curvature is non-vanishing, 
indicating the presence of thermodynamic 
interaction, and free of singularities,
indicating the absence of phase transitions. Similar results 
are obtained for generalizations of the BTZ black hole which 
include a Chern-Simons term and a dilatonic field. Small
logarithmic corrections of the entropy turn out to be represented
by small corrections of the thermodynamic curvature, reinforcing the
idea that thermodynamic curvature is a measure of thermodynamic 
interaction.

\end{abstract}

\maketitle
%\begin{keyword}
% Equilibrium thermodynamics \sep contact structure \sep Riemannian structure \sep 
% Weinhold's metric
%\end{keyword}

% main text
\section{Introduction}
\label{sec:int}
% Differential geometry is a very important tool of modern science, 
% specially of mathematical physics and its applications in physics,
% chemistry and engineering. 

The spacetime of a black
hole in (2+1) dimensions with negative cosmological constant
provides an example of a lower-dimensional toy model which shares
many of the important conceptual issues of general relativity in (3+1)
dimensions, but avoids some of the difficulties found in
mathematical computations. This spacetime is known as the
Ba\~nados-Teitelboim-Zanelli (BTZ) black hole \cite{btz},
and it warrants
attention in its own right (for a review, see \cite{carlip}).
A key feature of this model lies in the simplicity of
its construction. It is a spacetime with constant negative curvature
and is obtained as a discrete quotient of three-dimensional anti-de Sitter
space \cite{bhtz}. The BTZ spacetime is free of curvature
singularities. Even so, all characteristic features of black holes
such as the event horizon and Hawking radiation are present so that
this model is a genuine black hole. Furthermore, despite its simplicity,
the BTZ black hole plays an outstanding role in many of the recent
developments in string theory, specially in the context of the 
AdS/CFT conjecture \cite{btzstrings}. One of the most interesting
aspects of black holes is related to their thermodynamic properties. In
the case of the BTZ black hole, the extensive thermodynamic variables are the mass 
$M$, angular momentum $J$, and entropy $S$ which is proportional to the horizon area. 
The intensive variables are the angular velocity at the horizon $\Omega$ and 
the Hawking temperature $T$. Although these quantities satisfy the laws of macroscopic 
thermodynamics, their microscopic origin remains obscure, and it is believed that it
is related to the problem of quantization of gravity. 

On the other hand, it is possible to introduce differential geometric concepts
in ordinary thermodynamics. The most known structures were postulated by 
Weinhold \cite{wei1} and Ruppeiner \cite{rup79,rup95} who introduced Riemannian 
metrics in the space of equilibrium states of a thermodynamic system. These 
geometric structures can obviously be applied in black hole thermodynamics. For
instance, the components of Weinhold's metric are simply defined as the second 
derivatives of the mass with respect to the extensive variables. The calculations 
are straightforward, but the geometric properties of the 
resulting manifolds are puzzling \cite{aman03,aman06a}.
For instance, in the case of the BTZ black 
hole thermodynamics, where $M=M(S,J)$, the curvature of the equilibrium space turns
out to be flat \cite{caicho99,sarkar06,med08}. This flatness is usually interpreted
 as a consequence of the lack of thermodynamic interaction. However, if one applies a 
Legendre transformation $M\rightarrow \tilde M = M - J\Omega$, the resulting manifold
is curved. This result is not in agreement with ordinary thermodynamics which is manifestly
Legendre invariant. To overcome this inconsistency, the theory of geometrothermodynamics
(GTD) was proposed recently \cite{quev07,quev08,quevaz}. 
It incorporates arbitrary Legendre transformations \cite{arnold}
into the geometric 
structure of the equilibrium space in an invariant manner. In this work, we study
the equilibrium space of the BTZ black hole, and propose a thermodynamic metric whose 
curvature is non-zero and reproduces its main thermodynamic properties. In particular, we
will see that the thermodynamic curvature is free of singularities. In GTD, this 
is interpreted as a consequence of the non-existence of singular points at the level 
of the heat capacity, indicating that no (second order) phase transitions occur. 
It turns out that these results coincide with the predictions of ordinary black 
hole thermodynamics as proposed by Davies \cite{davies}. We also include in our analysis
the case of an additional Chern-Simons term and a dilatonic field at the level of 
the action. In all the cases presented in this work, it turns out that GTD correctly reproduces 
the thermodynamic properties of the corresponding system. Moreover, we analyze
the leading logarithmic corrections to the entropy and show that they correspond to
small perturbations at the level of the curvature of the equilibrium space. This can
be considered as an additional indication that the thermodynamic curvature can be 
used to measure the thermodynamic interaction of a system.

This paper is organized as follows. In Section \ref{sec:btz} we 
review the most important aspects of the BTZ black hole, emphasizing
the thermodynamic interpretation of its physical parameters. 
In Section \ref{sec:gtd} we use the formalism of
GTD to construct the thermodynamic
phase space and the space of equilibrium states for the BTZ black hole. 
Section \ref{sec:gen} is devoted to study the GTD of generalizations
of the BTZ black hole, including an additional  Chern-Simons  charge and
a dilatonic field. In all the cases we investigate the influence of small
corrections of the entropy on the thermodynamic curvature. 
 Finally, Section \ref{sec:con} contains 
discussions of our results and suggestions for further research.
Throughout this paper we use units in which $c=k_{_B}=\hbar =8G=1$.

%%%%%%%%%%%%%%%%%%%%%%%%%%%%%%%%%%%%%%%%%%%
\section{The BTZ black hole}
\label{sec:btz}

The BTZ black hole metric in spherical coordinates can be written as
\be
ds^2 = -  \left( - M + \frac{r^2}{{ l}^2} + \frac{J^2}{4 r^2}\right) d  t ^2 
+ r^2 \left(d\varphi -  \frac{J}{2r^2} dt\right)^2
+   \frac {dr^2}{- M + \frac{r^2}{{ l}^2} + \frac{J^2}{4 r^2}}
 \ ,
\label{btz}
\ee
where $M$ and $J$ are the mass and angular momentum, respectively.
The BTZ metric is a classical solution
of the field equations of (2+1)-gravity
which follow from the action
$ 
I = {1}/({16\pi })\int d^3x \sqrt{-g} (R-2\Lambda) \ ,
$
where $\Lambda= -1/{ l}^2$ is the cosmological constant.
The BTZ metric
is characterized by a constant negative curvature and, therefore, can be obtained as a
region of  anti-de Sitter space with an appropriate identification of the
boundaries \cite{btz}. The roots of the lapse function ($g_{tt}=0)$ 
\be
r_\pm^2 =
\frac{{ l}^2}{2}\left[ M \pm
\left(M^2 -\frac{J^2}{{ l}^2}\right)^{1/2}\right]
\ee
define the horizons $r=r_\pm$ of the spacetime. In particular, the null hypersurface
$r=r_+$ can be shown to correspond to an event horizon,
which in this case is also a Killing horizon, whereas the inner horizon at $r_-$
is a Cauchy horizon. 
From the expressions for the horizon radii the following useful relations are obtained
\be
M = \frac{r_+^2+r_-^2}{l^2} \ ,\qquad J=\frac{2r_+r_-}{l} \ .
\label{mj}
\ee
From the area-entropy relationship, $S = 4\pi r_+$, we obtain 
an expression of the form $S=S(M,J)$ that can be rewritten as
\be
\label{feq}
M = \frac{S^2}{16\pi^2l^2} + \frac{4\pi^2J^2}{S^2}\ .
\ee
This equation relates all the thermodynamic variables entering the BTZ metric in
the form $M=M(S,J)$ so that if we impose the first law of thermodynamics
$
dM = T d S + \Omega d J
$, 
the expressions for the temperature and the angular velocity can easily 
be computed as $T = \partial M/\partial S$, $\Omega = \partial M/\partial J$. 
It is convenient to write the final results in terms of the horizon radii 
by using the relations (\ref{mj}):
\be
T = \frac{r_+^2-r_-^2}{2\pi l^2 r_+}\ ,\qquad \Omega = \frac{r_-}{lr_+} \ . 
\ee
The temperature is always positive and vanishes only in the case of an extremal black hole,
i. e., when $r_+=r_-$. The heat capacity at constant 
values of $J$ is given as 
\be
C= T\frac{\partial T}{\partial S} = \frac{4\pi r_+ (r_+^2-r_-^2)}{r_+^2 + 3 r_-^2}\ .
\ee
Following the fundamentals of black hole thermodynamics as formulated by Davies 
\cite{davies}, the main thermodynamic properties of the BTZ black hole can be derived 
from the behavior of its thermodynamic variables $M$, $T$, $\Omega$ and $C$ in terms
of the extensive variables $S$ and $J$. We see that all thermodynamic variables are 
well-behaved, except perhaps in the extremal limit $r_+ = r_-$, where the Hawking
temperature and the heat capacity vanish. Since an absolute zero temperature is not allowed
by the third law of thermodynamics, we conclude that the thermodynamic description breaks
down in the extremal limit. 
The fact that the heat capacity $C$ is always positive and free of singular points is 
usually interpreted as an indication that the BTZ black hole is a thermodynamically stable 
configuration where no phase transitions can occur.  
This is in contrast with black hole configurations in higher dimensions which, in 
general, are characterized by regions of high instabilities and a rich phase transitions 
structure. 

It should be mentioned that Davies' formulation of phase transitions for black holes
is not definitely settled and is still a subject of discussion. Alternative 
criteria for the existence of phase transitions of black holes 
have been proposed in different contexts \cite{cur81,pav91,kok93,ook93}. A definite definition could be formulated only on the basis of a microscopic description that 
would lead to the ordinary macroscopic thermodynamics of black holes in the appropriate 
limit. Such a microscopic model for black holes must be related to a hypothetical model 
of quantum gravity which is still out of reach. We therefore use the intuitive 
definition of phase transitions as it is known from ordinary thermodynamics of black holes
\cite{davies}.

%%%%%%%%%%%%%%%%%%%%%%%%%%%%%%%
%%%%%%%%%%%%%%%%%%%%%%%%%%%%%%%%
\section{Geometrothermodynamics of the BTZ black hole}
\label{sec:gtd}

For the geometric description of the thermodynamics of the BTZ black
hole in GTD, we first introduce the 5-dimensional 
phase space ${\cal T}$ with coordinates $\{M,S,J,T,\Omega\}$, a contact 1-form
$\Theta = dM - TdS - \Omega dJ$, and an invariant metric
\be
G = (dM - T dS - \Omega dJ)^2 + (TS + \Omega J) \left( - dT d S + d\Omega dJ\right)\ .
\label{gup}
\ee
Similar metrics were obtained in GTD in order to propose an invariant geometric 
description of the thermodynamics of higher dimensional black holes \cite{aqs08,qs08}.
The triplet $({\cal T},\Theta,G)$ defines a contact Riemannian manifold that plays 
an auxiliary role in GTD. It is used to properly handle the invariance with respect 
to Legendre transformations. In fact, a Legendre transformation involves in general 
all the thermodynamic variables $M,S,J,T$, and $\Omega$ so that they must be independent
from each other as they are in the phase space. We introduce also the geometric structure
of the space of equilibrium states ${\cal E}$ in the following manner:  ${\cal E}$ is
 a 2-dimensional submanifold of ${\cal T}$ that is defined by the smooth embedding map
$\varphi: {\cal E} \longrightarrow {\cal T}$,
satisfying the condition that 
the ``projection" of the contact form $\Theta$ on ${\cal E}$ vanishes, i. e., 
$\varphi^* (\Theta) = 0$, where $\varphi^*$ is the pullback of $\varphi$, and
that $G$ induces
a Legendre invariant metric $g$  on ${\cal E}$ by means of $g=\varphi^*(G)$.
 In principle, any 2-dimensional subset of
the set of coordinates of ${\cal T}$ can be used to coordinatize ${\cal E}$. For
the sake of simplicity, we will use
the set of extensive variables $S$ and $J$ which in ordinary thermodynamics corresponds 
to the energy representation. Then, the embedding map for this specific choice is 
 $\varphi: \{S, J\} \longmapsto \{M(S,J),S,J,T(S,J),\Omega(S,J)\}$. 
The condition
$\varphi^*(\Theta)=0$ is equivalent to the first law of thermodynamics and the 
conditions of thermodynamic equilibrium 
\be
dM = T dS + \Omega d J \ , \quad  T = \frac{\partial M}{\partial S}\ ,\quad
\Omega = \frac{\partial M}{\partial J}\ ,
\ee
whereas the induced metric becomes
\be
\label{gdown}
g = \left(S\frac{\partial M}{\partial S} + J\frac{\partial M}{\partial J}\right)
\left(-\frac{\partial^2 M}{\partial S^2} d S^2 
+ \frac{\partial^2 M}{\partial J^2} d J^2 \right)\ .
\ee
This metric determines all the geometric properties of the equilibrium space ${\cal E}$.
We see that in order to obtain the explicit form of the metric it is only necessary to 
specify the thermodynamic potential $M$ as a function of $S$ and $J$. In ordinary 
thermodynamics this function is usually referred to as the fundamental equation from 
which all the equations of state can be derived \cite{callen}.

In general, it is possible to show that any metric  in ${\cal E}$ can be obtained as the
pullback of a metric in ${\cal T}$. In particular, the Weinhold metric $g^W=
(\partial^2 M/\partial E^a \partial E^b )d E^a d E^b $, with $E^a =\{S,J\}$, 
can be shown to be generated by a metric of the form 
$ G^W = dM^2 -(TdS + \Omega dJ)^2 + dS dT + d\Omega d J $ which can be shown to be
non invariant with respect to Legendre transformations \cite{quev07}. 
This explains why Weinhold's metric
leads to contradictory results when different thermodynamic potentials are used in its
definition. 
 
\begin{figure}
%\begin{center}
\includegraphics[width=7cm]{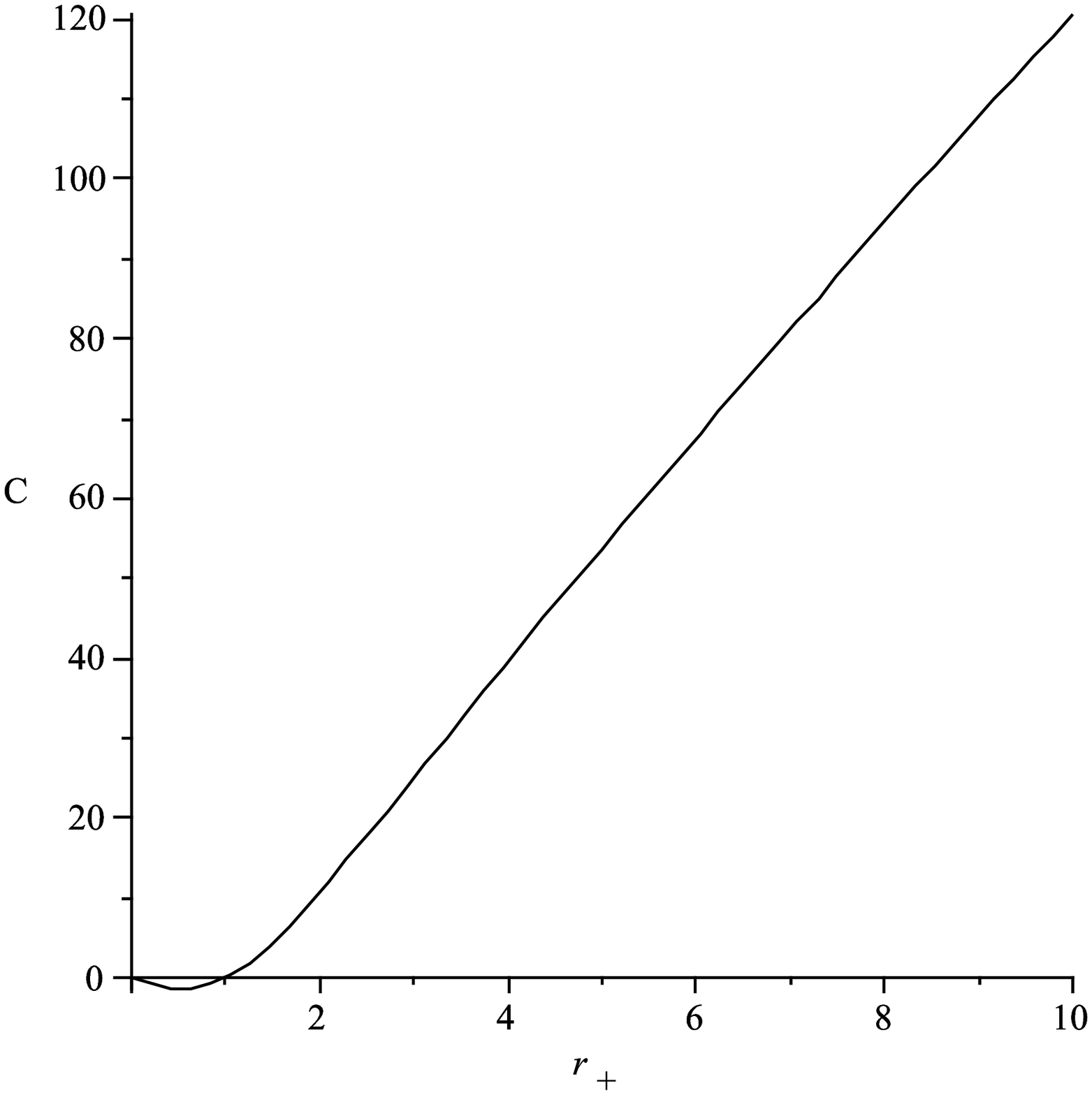}
\includegraphics[width=7cm]{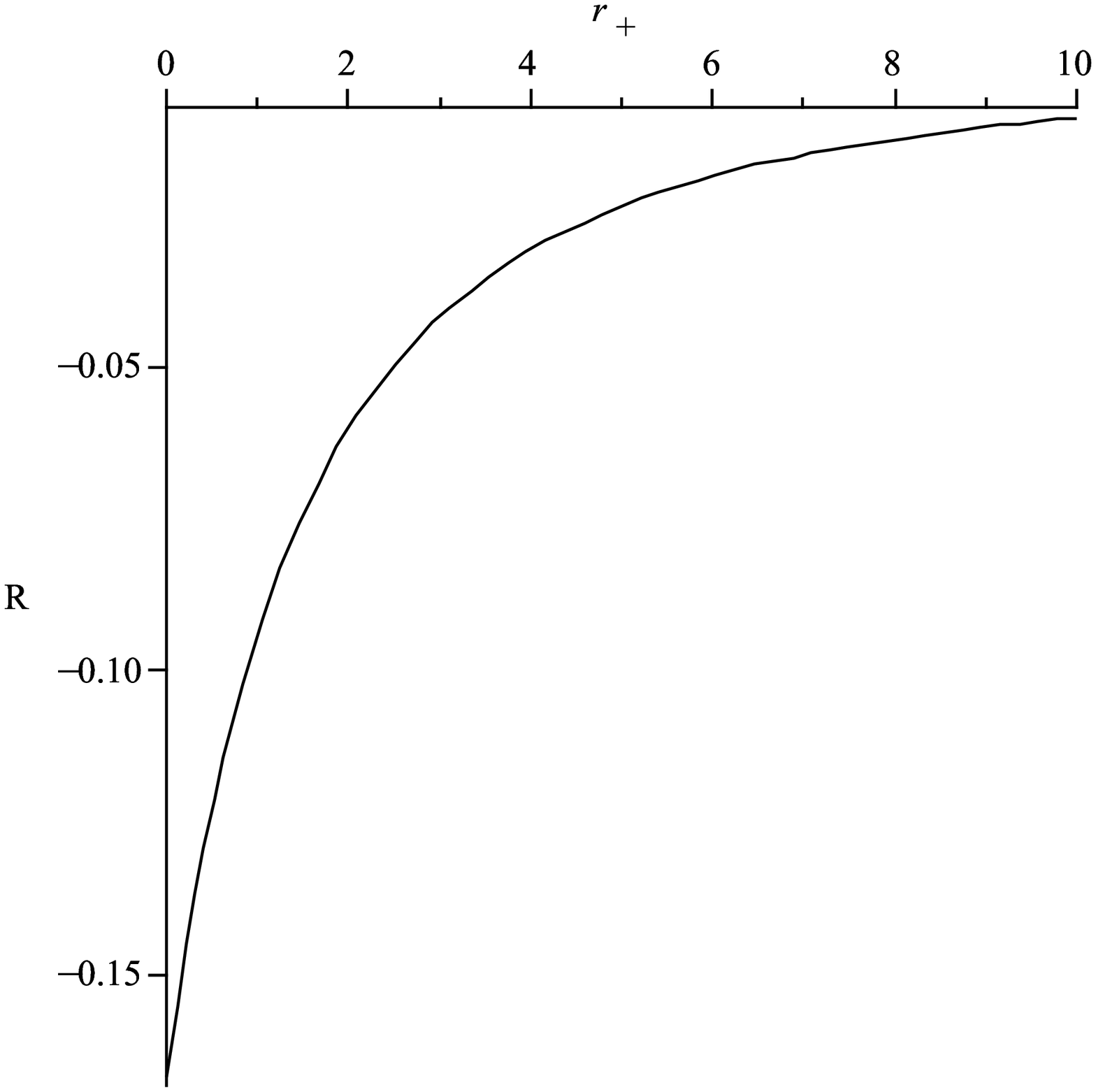}
%\end{center}
\caption{Heat capacity and thermodynamic curvature of the BTZ black hole. 
A typical behavior is depicted
for the specific values $l=1$ and $r_-=1$. The curvature is completely regular for the 
entire domain of the horizon radii. }
\label{fig:fig1}
\end{figure}

As for the BTZ black hole, the fundamental equation $M=M(S,J)$ follows from the area-entropy 
relationship and is given in Eq.(\ref{feq}). Then, it is easy to compute the explicit 
form of the thermodynamic metric (\ref{gdown}) which, using the expressions for $S$ and $J$ 
in terms of $r_+$ and $r_-$, can be written as
\be
\label{gbtz}
g= - \frac{r_+^2 + 3 r_-^2}{4\pi^2 l^4} dS^2 + \frac {1}{l^2} dJ^2 \ .
\ee
The corresponding thermodynamic curvature turns out to be non zero and the scalar curvature 
can be expressed as
\be
R= -\frac{3}{2}\frac{l^4}{(r_+^2 + 3 r_-^2)^2} \ .
\ee
The general behavior of this curvature is illustrated in figure \ref{fig:fig1}. The 
equilibrium manifold is a space of negative curvature for any values of the horizon 
radii and no singular points are present. This means that thermodynamic interaction
is always present and that no phase transitions can take place. Consequently, the 
BTZ black hole corresponds to a stable thermodynamic configuration. This interpretation
coincides with the interpretation derived in Section \ref{sec:btz} from the 
analysis of the thermodynamic variables. The heat capacity vanishes at the extremal limit
$r_+=r_-$ and becomes negative for $r_+<r_-$. This region is, however, not allowed by the
definition of the horizon radii. 
We conclude that the geometry of the equilibrium
space correctly describes the thermodynamic behavior of the BTZ black hole. 
Indeed, one of the main goals of GTD is to interpret thermodynamic curvature as a measure of
thermodynamic interaction and curvature singularities as points of phase transitions. We see
that the thermodynamic metric proposed in GTD for the BTZ black hole meets these goals. 

We also 
see from figure \ref{fig:fig1} that the thermodynamic curvature is regular even at the extremal 
limit and below. 
This is probably a consequence of the fact the there is no singularity inside the 
horizon. In fact, in the 4-dimensional case the geometric description of black hole thermodynamics 
breaks down in the region $r_->r_+$  as a consequence of the presence of a naked singularity \cite{aqs08}. 

The above analysis in terms of the thermodynamic potential $M=M(S,J)$ is usually interpreted
as corresponding to the canonical ensemble \cite{cal00}. In GTD, it is possible to consider
different ensembles at the geometric level. In fact, in ordinary thermodynamics 
different ensembles are related by Legendre transformations. Then, 
the auxiliary phase space ${\cal T}$ is the appropriate arena to handle different
statistical ensembles. Consider, for instance, the grand-canonical ensemble whose
thermodynamic potential is the Gibbs potential $\tilde M$ which can be introduced 
in ${\cal T}$ by the Legendre transformation $\{M,S,J,T,\Omega\}\longrightarrow 
\{\tilde M, \tilde S, \tilde J, \tilde T, \tilde\Omega\}$ with
$M = \tilde M - \tilde S \tilde T - \tilde J \tilde \Omega$ and 
$S = - \tilde T,\ J = -\tilde \Omega,\ T= \tilde S, \ \Omega = \tilde J$. 
This transformation leaves the metric $G$ invariant in the sense that in the new coordinates 
it can be written as
\be
G = (d\tilde M - \tilde T d\tilde S - \tilde\Omega d\tilde J)^2 + 
(\tilde T\tilde S +\tilde \Omega \tilde J) \left( - d\tilde T d \tilde S + d\tilde\Omega d\tilde J\right)\ .
\label{gupnew}
\ee
Then, it is clear that the induced metric $g$ in ${\cal E}$ is also invariant in the same
sense. Consequently, the geometry of the equilibrium space can also be applied in 
a straightforward manner to the grand-canonical ensemble. 

The choice of the metric $G$ as given in Eq.(\ref{gup}) is now clear. All the terms are of the
form $E^a I^a, a=1,2,$ where $E^a=(S,J)$ are the extensive variables and $I^a=(T,\Omega)$ are 
the intensive variables. This choice guarantees invariance with respect to Legendre transformations
which interchange extensive and intensive variables. We will use this criterion to construct analogous metrics in Section \ref{sec:gen}.  

To finish this section, it is worth mentioning that 
it is possible to consider the cosmological constant $\Lambda$ as an additional extensive 
thermodynamic variable. In this case, the equilibrium space becomes 3-dimensional and it turns out that instead of $\Lambda$ it is necessary to consider the radius of curvature
$l^2$ as the additional variable. However, it would be necessary to perform a more detailed
analysis in terms of statistical ensembles in order to understand the radius of curvature 
as a realistic thermodynamic variable \cite{selene}.

%%%%%%%%%%%%%%%%%%%%%%%%%%%%%%%%%%%%%%%%%%%%%%%%%%%%%
%%%%%%%%%%%%%%%%%%%%%%%%%%%%%%%%%%%%%%%%%%%%%%%%%%%%%%
\section{Geometrothermodynamics of BTZ generalizations}
\label{sec:gen}
In this section we will investigate the geometry of the equilibrium space of
certain generalizations of the BTZ black hole. Our goal is to see whether 
these generalizations can also be interpreted as a sources of thermodynamic
interaction in the sense that they affect the thermodynamic curvature of the
equilibrium space. We will focus our analysis on BTZ black holes with an additional 
Chern-Simons (CS) charge and a dilatonic field as well as thermal fluctuations 
of the BTZ black hole. We will see that in all the cases GTD correctly describes
the thermodynamics of the corresponding system.

\subsection{The Chern-Simons charge}
\label{sec:cs}
The inclusion of CS charges is important in the study of gravitational 
anomalies and for the case of (2+1)-gravity it was performed in \cite{kralar06}
and \cite{sol06}. The addition of the gravitational CS-term to the Einstein-Hilbert
action with cosmological constant results in a new theory that is known as 
topologically massive gravity \cite{djt82}. The BTZ solution turns out to be an exact  solution
of the corresponding field equations with a different mass and angular momentum \cite{sol06}
\be
M = M_0 - \frac{k}{l^2} J_0     \ ,\quad J = J_0 - k M_0\ ,
\ee
where $M_0$ and $J_0$ are the mass and angular momentum parameters of the original 
BTZ solution as given in Eq.(\ref{mj}), and $k$ is the Chern-Simons coupling constant.
Moreover, the expression for the entropy results
modified into \cite{sol06}
\be
\label{entcs}
S= 4\pi\left( r_+ -\frac{k}{l}r_-\right) \ .
\ee
It is then easy to show that in terms of the new parameters $M$ and $J$, the horizon radii
can be expressed as
\be
r_\pm = \frac{l}{2}
\left[ \left( \frac{lM+J}{l-k} \right)^{1/2} \pm 
\left(\frac{lM-J}{l+k}\right)^{1/2}\right]\ ,
\ee
which can then be introduced into the modified entropy (\ref{entcs}) 
to obtain the fundamental 
equation in the entropy representation $S=S(M,J)$. The resulting equation can then be rewritten as 
\be
\label{febtzcs}
M= \frac{1}{8\pi^2k^2}\left[
S^2+8\pi^2kJ + \frac{S}{l}\sqrt{(l^2-k^2)(S^2+16\pi^2kJ)}\right]\ ,
\ee
which is the fundamental equation in the mass representation. Introducing this 
fundamental equation into Eq.(\ref{gdown}) and expressing the result in terms of the the 
horizon radii, we obtain the thermodynamic metric 
\be
\label{gbtzcs}
g= \frac{2lr_+^3+k r_- (r_-^2-3r_+^2)}{2l^2r_+^4(l^2-k^2)}\left[
\frac{kr_-(r_-^2+3r_+^2)-lr_+(r_+^2+3r_-^2)}{4\pi^2l^2} dS^2
+(lr_+-kr_-)dJ^2\right]\ ,
\ee
which describes the geometry of the equilibrium space and reduces to the thermodynamic
metric for the BTZ black hole (\ref{gbtz}) in the limiting case $k\rightarrow 0$.
 Other important thermodynamic quantities 
can be computed from the fundamental equation (\ref{febtzcs}). For instance, the Hawking
temperature $T=\partial M/\partial S$ and the heat capacity 
$C= T (\partial^2 M /\partial S^2)^{-1}$ at constant $J$ 
can be written in the form
\be
\label{tcbtzcs}
T= \frac{r_+^2 - r_-^2}{2\pi l^2 r_+}\ ,\quad
C= \frac{4\pi(l^2-k^2) r_+^2 (r_+^2-r_-^2)}{l[ lr_+(r_+^2+3r_-^2) - kr_-(r_-^2+3r_+^2)]} \ .
\ee
\begin{figure}
%\begin{center}
\includegraphics[width=7cm]{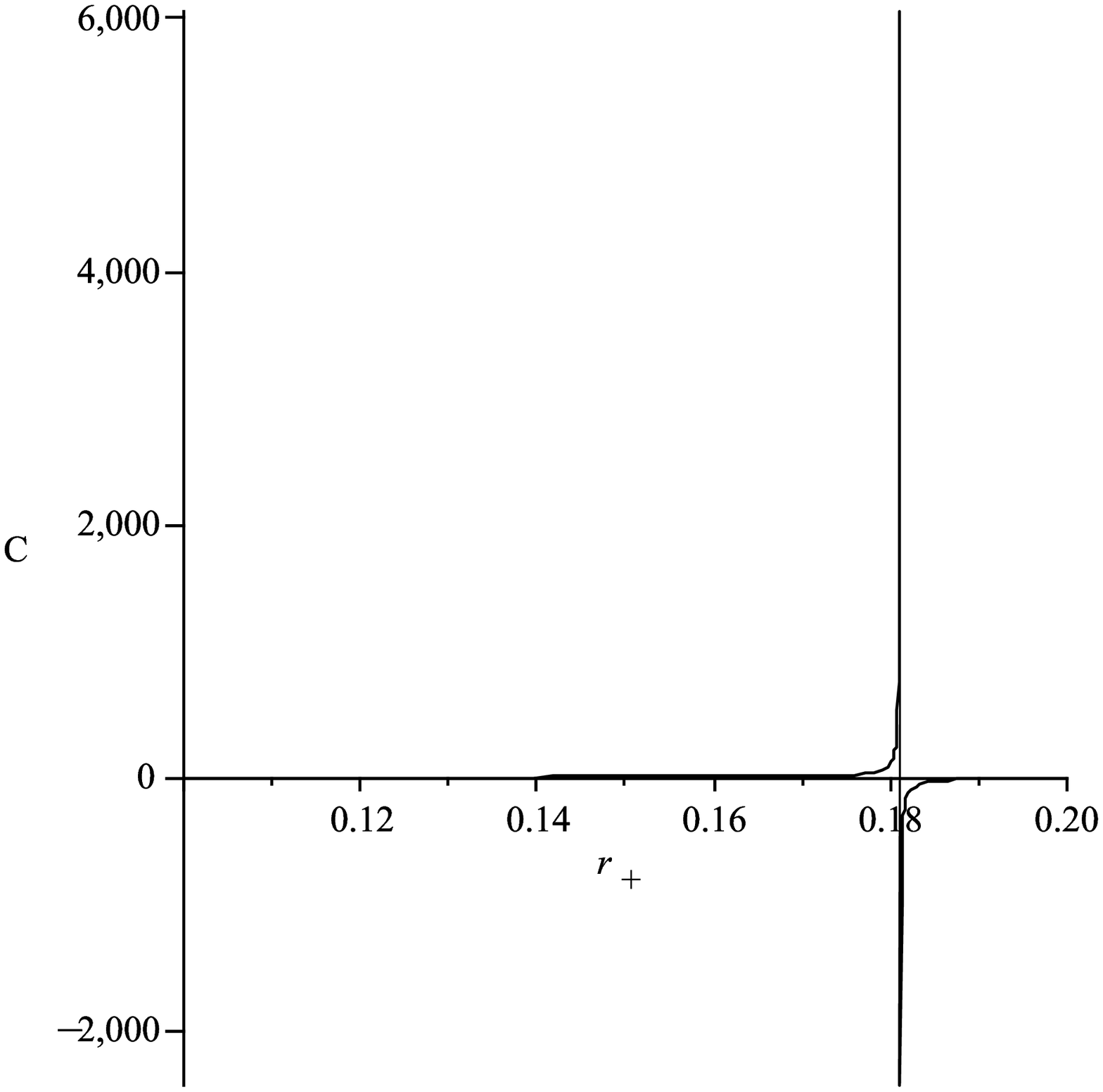}
\includegraphics[width=7cm]{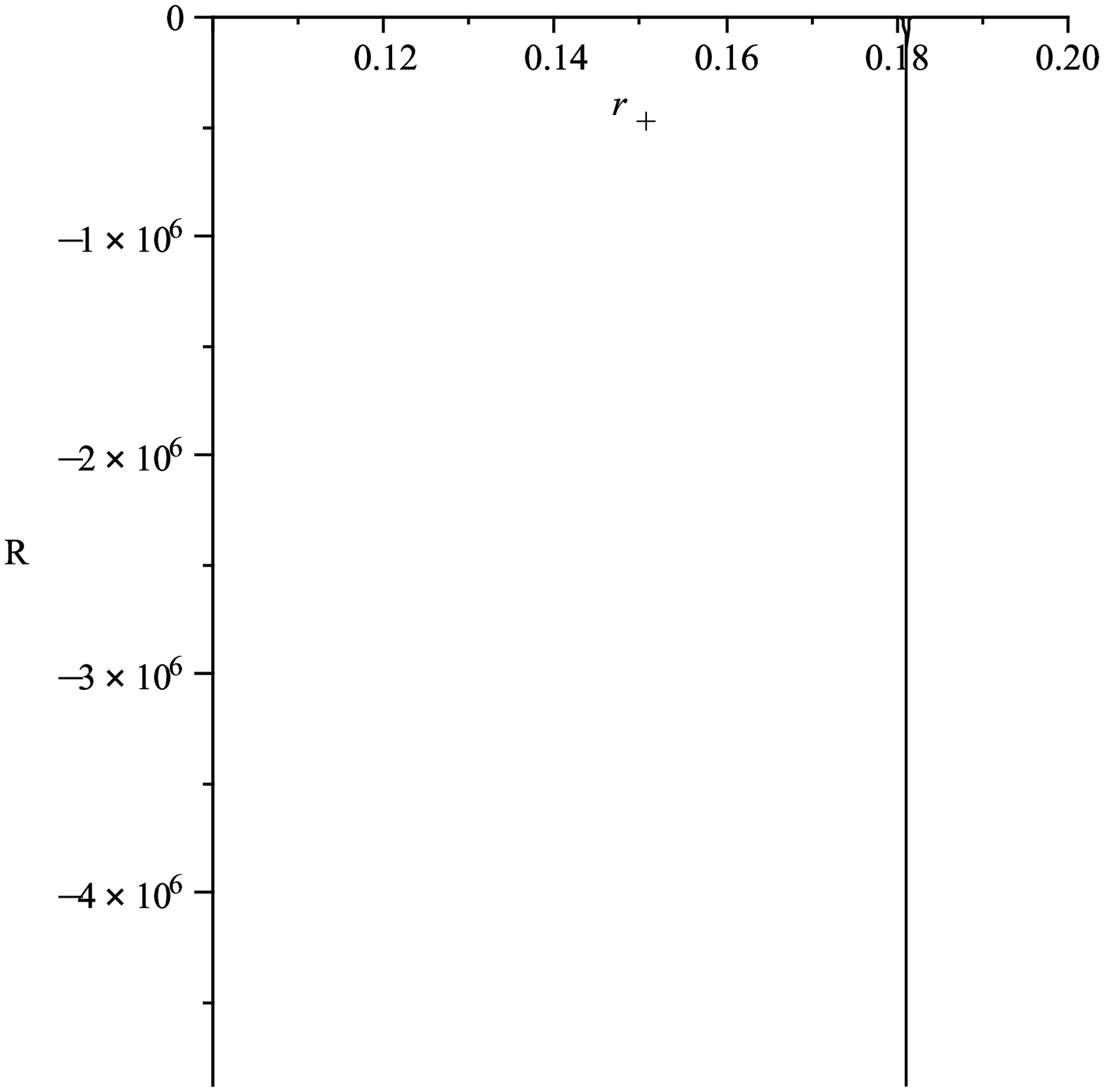}
%\end{center}
\caption{The heat capacity and the thermodynamic curvature 
of the BTZ-CS black hole. A typical behavior is depicted
for the specific values $l=1$, $k=1/2$ and $r_-=1$. There is only one singular point
at $r_+\approx 0.18$, indicating a possible phase transition.}
\label{fig:btzcsc}
\end{figure}

The computation of the scalar curvature of the thermodynamic metric (\ref{gbtzcs}) is 
straightforward, but the resulting expression cannot be written in a compact form. 
We analyzed numerically the behavior of the thermodynamic curvature and found that 
it has a singular point which coincides with the singular point of the heat capacity 
(\ref{tcbtzcs}). This behavior is illustrated in figure \ref{fig:btzcsc}. However, as can 
be seen in the graphic, the singularity is situated at a point $r_+<r_-$ which contradicts
the physical significance of the outer $r_+$ and inner $r_-$ horizon radii. Moreover,
the Hawking temperature becomes negative for values $r_+<r_-$.

\begin{figure}
%\begin{center}
\includegraphics[width=7cm]{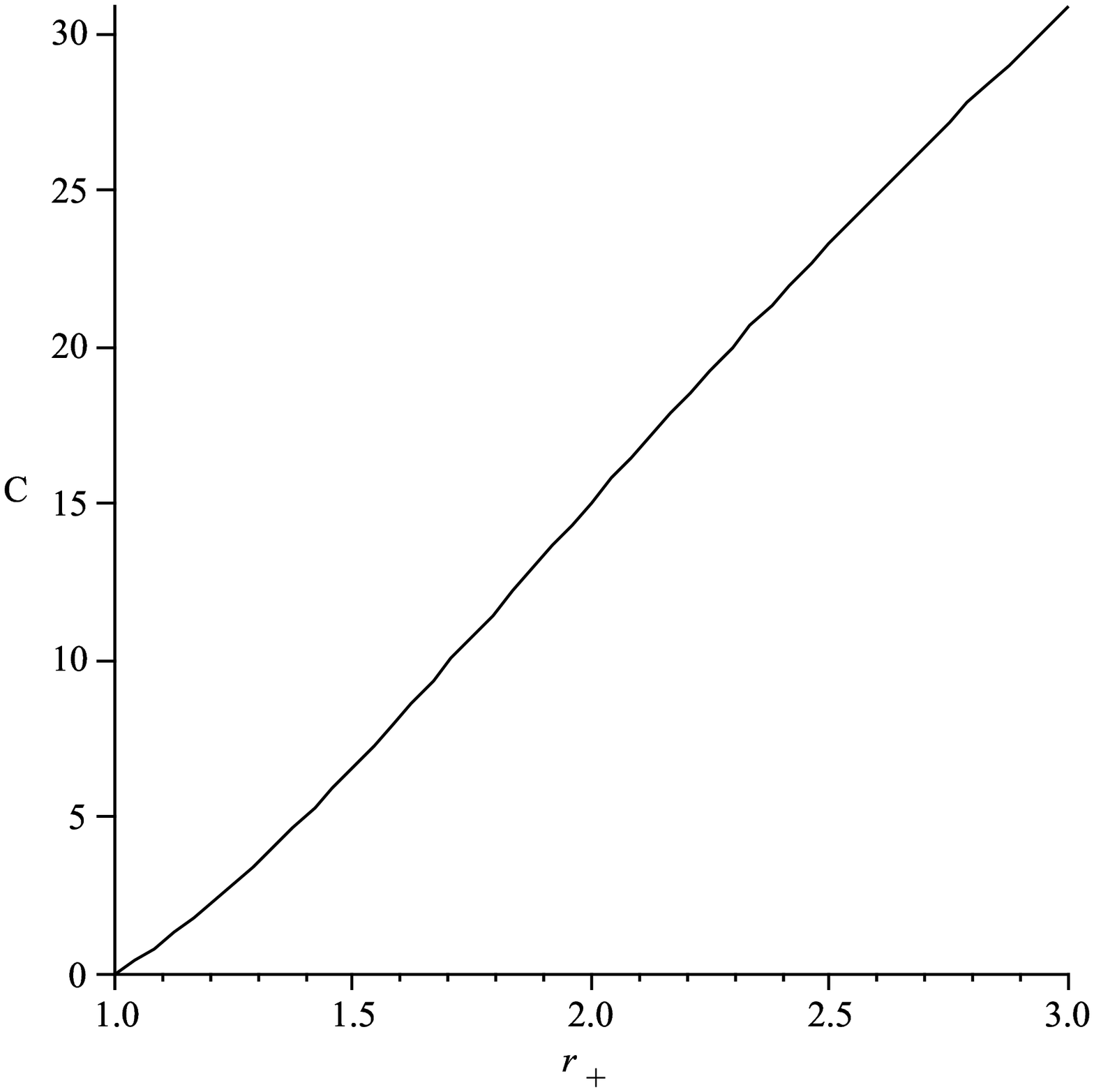}
\includegraphics[width=7cm]{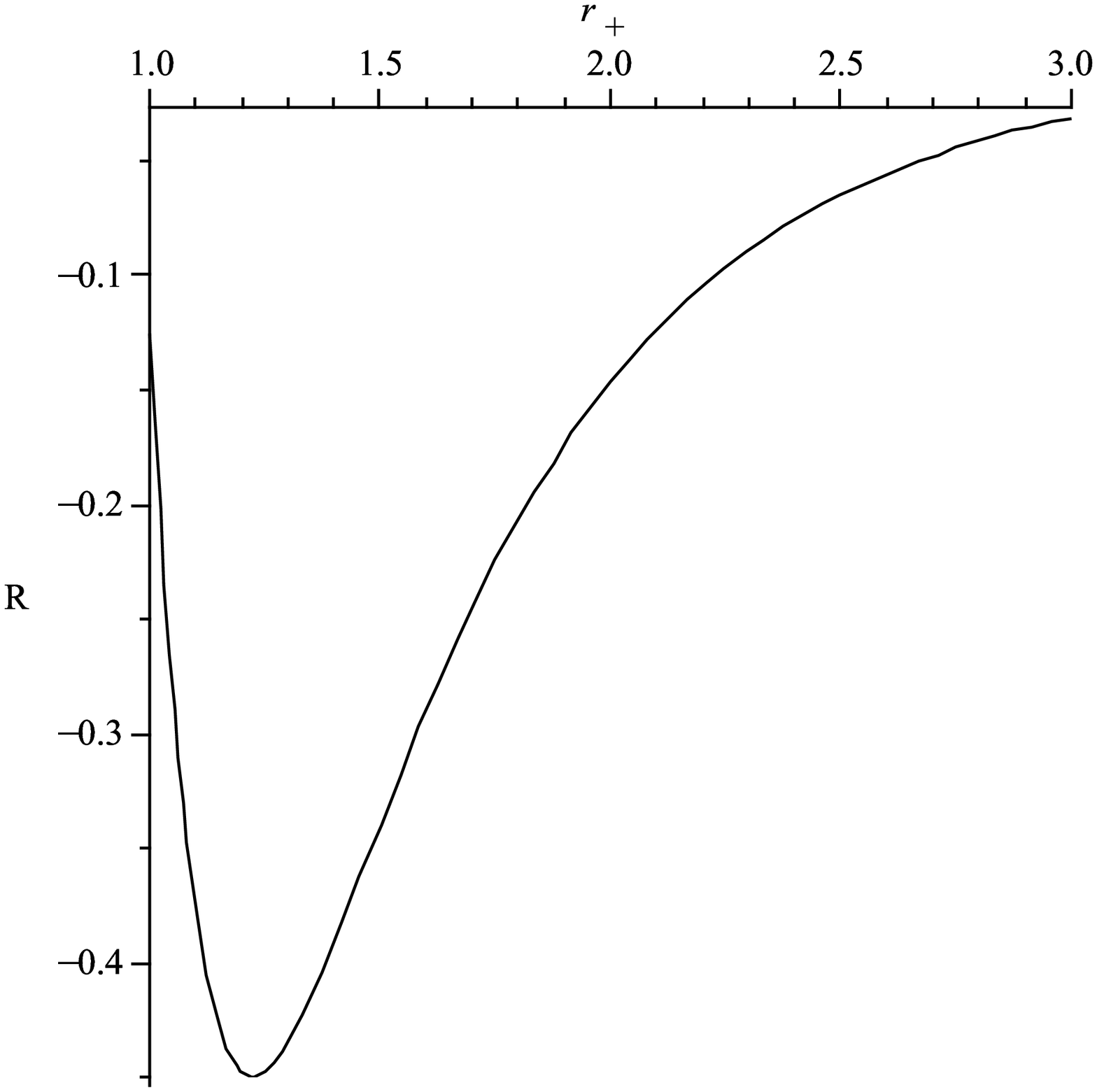}
%\end{center}
\caption{The heat capacity and the thermodynamic curvature 
of the BTZ-CS black hole for the choice of values 
$l=1$, $k=1/2$ and $r_-=1$ and the interval $r_+> r_-$. 
The heat capacity is always positive and the thermodynamic curvature 
is free of singularities, ensuring thermodynamic stability.}
\label{fig:btzcs1}
\end{figure}

In the physically meaningful interval $r_+\geq r_-$ with the additional condition 
$l\geq k$, which is necessary in order to the fundamental equation 
(\ref{febtzcs}) to be well defined, it can be shown that the heat capacity is 
always positive, indicating that the BTZ-CS black hole is thermodynamically stable. 
The thermodynamic curvature corresponding to the metric (\ref{gbtzcs}) in this
interval  is as depicted in figure \ref{fig:btzcs1}.  
The heat capacity and the thermodynamic curvature 
are indeed affected by the presence of the CS charge, but the general behavior 
remains the same. We only observe that the thermodynamic curvature has now 
a minimum value from which it grows as the outer horizon radius approaches the inner 
radius. Nevertheless, the behavior is completely regular in the entire physical interval.
The fact that the curvature
is free of singularities is interpreted in GTD as an indication of the absence 
of phase transitions structure.  An additional 
numerical analysis of the thermodynamic curvature 
shows that it diverges in the limit $k\rightarrow l$. 
This is in agreement with 
the fundamental equation (\ref{febtzcs}) which requires that $k\leq l$.
We conclude that the geometry of the equilibrium
space, as described by the thermodynamic metric (\ref{gbtzcs}), correctly describes
the thermodynamics of the BTZ-CS black hole.

\subsection{The dilatonic field}
\label{sec:dil}
Scalar fields are believed to play an important role in modern physics. 
Nearly all generalized gravity 
theories such 
as scalar-tensor and Kaluza-Klein theories involve at least one scalar field.  Moreover, 
a specific scalar field arises from the string theory - the so-called dilatonic field. 
This is one of the reasons why during the last two decades many works have been devoted to the study
of the dilatonic field in different scenarios. In (2+1) gravity, for instance, the influence of a dilaton
on the BTZ black hole was investigated in \cite{chan97} where solutions were found to the field equations 
that follow from the gravitational action  coupled to a self-interacting dilatonic field. Since 
the dilatonic potential can take, in principle, any desired form, the number of possible solutions could 
be quite big. We choose a simple trigonometric potential which is given as a truncated series of $\cot\sqrt{2}\phi$
so that the 
total Lagrangian becomes \cite{chan97}
\be
{\cal L} = R - 2\Lambda - 4 (\triangle \phi)^2 - 2 \Lambda \cot^2\sqrt{2}\phi +
 \alpha \left(1+\cot^2\sqrt{2}\phi\right)\cot^4\sqrt{2}\phi \ ,
\ee
and a particular solution can be expressed as
\be
\label{btzdil}
ds^2=-\left( -M + \frac{M L}{r^2} - \Lambda r^2\right)dt^2 +
\frac{\left(1-\frac{2L}{r^2}\right)^2}{-M + \frac{M L}{r^2}-\Lambda r^2}dr^2 + r^2d\varphi^2\,,\quad
\phi = \frac{1}{\sqrt{2}} {\rm arcsec}\frac{r}{\sqrt{2L}}\ ,
\ee
where $L$ and $M$ are integration constants. Moreover, the coupling constant turns out to be fixed by
$\alpha = 2\Lambda + M /(2L)$. For the sake of concreteness we assume that $M>0$ and the form of the 
dilatonic field implies that $L>0$. This solution is a generalization of the BTZ black hole in the sense 
that in the asymptotic limit $r\rightarrow \infty$, the metric (\ref{btzdil}) transforms into the 
non-rotating $(J=0)$ BTZ metric (\ref{btz}). To our knowledge, there is no known dilatonic generalization 
for the rotating $(J\neq 0)$ case. Since $\Lambda <0$, the solution (\ref{btzdil}) describes a black hole
with outer and inner horizons located at 
\be
\label{horbtzdil}
r_\pm^ 2  = \frac{M\pm\sqrt{M^2-4|\Lambda| ML}}{ 2|\Lambda| }\ ,
\ee
so that the main parameters become
\be
\label{parbtzdil}
M= |\Lambda| (r_+^2+r_-^2)\ ,\quad L = \frac{r_+^2 r_-^2}{r_+^2+r_-^2}\ .
\ee
As before, the entropy of the black hole is given as $S=4 \pi r_+$ and its explicit expression can be rewritten
as 
\be
\label{febtzdil}
  M=\frac{|\Lambda| S^{4}}{16\pi^2(S^2-16\pi^2L ) }\,.
\ee
From the point of view GTD, this relationship represents the fundamental equation $M=M(S,L)$ from which all
the thermodynamic information can be obtained. This means that we will consider $L$ as a thermodynamic variable
and the first law of thermodynamics can be written as $dM = T dS + \psi d L$ where $\psi = \partial M /\partial L$ 
is the thermodynamic variable dual to $L$.  Then, using the relationships (\ref{parbtzdil}), 
the temperature and heat capacity at constant $L$ are given by
\be
T = \frac{|\Lambda| (r_+^4 - r_-^4)}{2\pi r_+^3}\ ,\quad 
C=  \frac{4\pi r_-^3(r_+^2-r_-^2)}{r_+^4 - r_+^2r_-^2 + 4r_-^4} \ .
\ee
We see that the temperature is always positive, except at the extremal black hole limit where it vanishes. 
The heat capacity is positive definite, indicating that the black holes is completely 
stable. Within the allowed
interval $r_+>r_-$, the denominator of the heat capacity is positive definite so that $C$ is regular everywhere.
In standard black hole thermodynamics this is interpreted as an indication that no phase transitions can occur. 

\begin{figure}
%\begin{center}
\includegraphics[width=7cm]{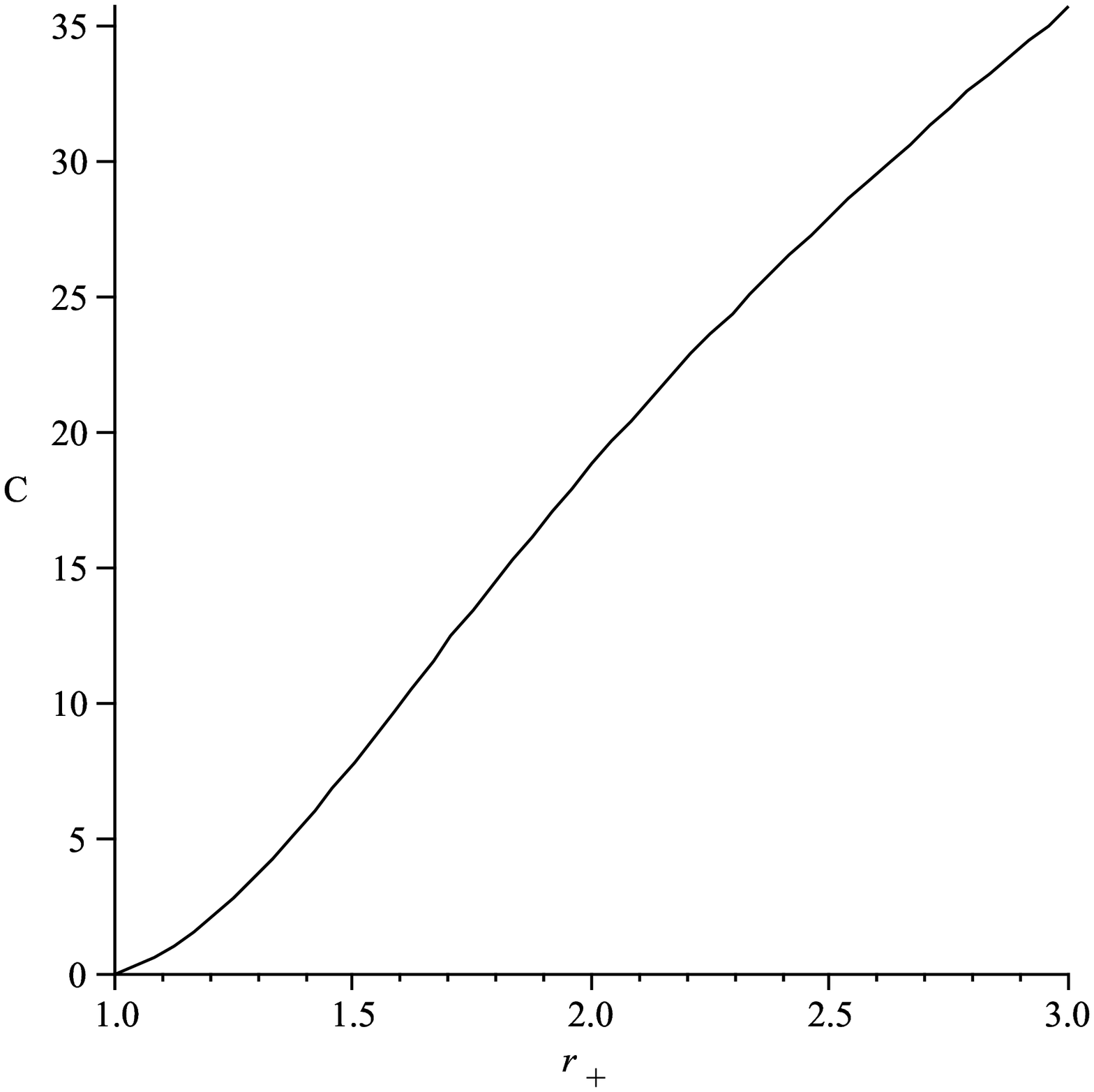}
\includegraphics[width=7cm]{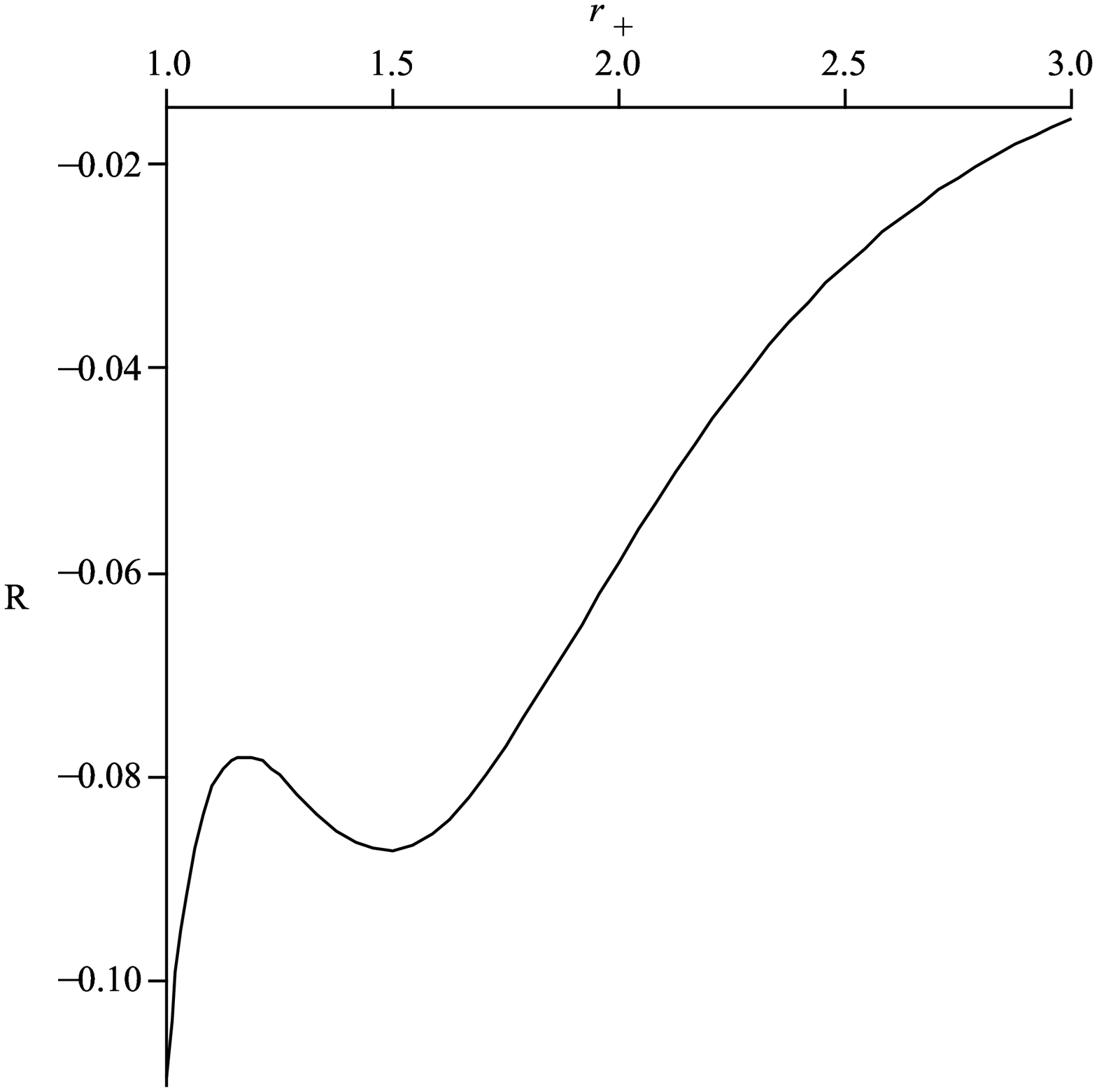}
%\end{center}
\caption{The heat capacity and the thermodynamic curvature 
of the BTZ black hole coupled to a dilatonic field. The parameter
values are $\Lambda=1$,  $r_-=1$ and $r_+> 1$. 
In this region, the heat capacity is always positive and the 
thermodynamic curvature is free of singularities.}
\label{fig:btzdil}
\end{figure}

We now turn to the geometric description of the thermodynamics of the black hole (\ref{btzdil}) in the context of GTD.
According to the fundamental equation (\ref{febtzdil}), the coordinates of the thermodynamic phase space ${\cal T}$
can be chosen as $\{M,S,L,T,\psi\}$ and the metric $G$ of ${\cal T}$ is similar to (\ref{gup}) with $J$ and $\Omega$ replaced
by $L$ and $\psi$, respectively. As the set of  coordinates for the equilibrium space ${\cal E}$ we choose $\{S,L\}$. Then,
the geometric construction of ${\cal E}$ is similar to the one presented in Section \ref{sec:gtd}, replacing
everywhere   $J$  by $L$ and  $\Omega$ by $\psi$. The final form of the metric $g$ of ${\cal E}$ is then
\be
\label{gdowndil}
g = \left(S\frac{\partial M}{\partial S} + L\frac{\partial M}{\partial L}\right)
\left(-\frac{\partial^2 M}{\partial S^2} d S^2 
+ \frac{\partial^2 M}{\partial L^2} d L^2 \right)\ .
\ee
The question now is whether this thermodynamic metric reproduces the thermodynamic behavior of the dilatonic BTZ black
hole (\ref{btzdil}) as described above. In GTD as in ordinary thermodynamics, the entire 
thermodynamic information can be extracted from the 
fundamental equation. Introducing the explicit form of the fundamental equation $M=M(S,L)$
as given in Eq.(\ref{febtzdil}) into the general metric $g$ given in Eq.(\ref{gdowndil}),
the resulting metric can be expressed in terms of the horizon radii as
\be
\label{gdownbtzdil}
ds^2 = \frac{2\Lambda^2(2r_+^2- r_-^2)(r_+^2+r_-^2)^2}{r_+^8} \left[ -\frac{r_+^4-r_+^2 r_-^2 + 4 r_-^4}{16\pi^2}dS^2
+ \frac{(r_+^2+r_-^2)^2}{r_+^2} d L^2 \right]
\ .
\ee
The corresponding scalar curvature is calculated in the standard way and we obtain
\be
\label{tcbtzdil}
R = -\frac{ r_+^6(40r_+^6 -69r_+^4r_-^2 + 39r_+^2r_-^4 + 4r_-^6)}
{4\Lambda^2(r_+^2+r_-^2)(2r_+^2-r_-^2)^3(r_+^4-r_+^2r_-^2+4r_-^4)^2} \ .
\ee
The general behavior of this curvature is illustrated in figure \ref{fig:btzdil}.
We see that the metric (\ref{gdownbtzdil}) describes a space of negative curvature
which, in the region $r_+> r_-$, is free of singularities. This behavior indicates
that no phase transitions can occur. The thermodynamic curvature (\ref{tcbtzdil}) 
describes a system which is in stable in thermodynamic equilibrium. This coincides with
the result of analyzing the heat capacity as described above. The general behavior 
of the heat capacity is also depicted in figure \ref{fig:btzdil}.

%%%%%%%%%%%%%%%%%%%%%%%%%%%%%%%%%%%%%%%%%%%%%%%%%%%%%%%%%%%%%%%%%%%%%%%%%%%%%
\subsection{Thermal fluctuations}
\label{sec:fluct}

If the canonical ensemble of a specific system is thermodynamically stable, 
it is well known that its  entropy is subject to logarithmic and polynomial corrections,
when thermal corrections are taken into account.  If $S_0$ denotes the 
entropy calculated in the canonical ensemble of a thermodynamic system with 
temperature $T$ and heat capacity $C$, the leading term of the thermal fluctuations leads to the entropy 
correction \cite{statistics}
\be
\label{corr0}
S= S_0 -\frac{1}{2} \ln (CT^2) \ .
\ee
This and higher order corrections have been analyzed for many classes of black holes \cite{das02,more05}. It has been pointed out that, up to an additive constant,
the leading correction is logarithmic
\be
\label{corr}
S= S_0 -\frac{3}{2} \ln (S_0) \ ,
\ee 
and of quite general nature in the sense that it can be derived from a semiclassical approach 
as well as from completely different 
approaches to quantum gravity \cite{carlip00}. 
The BTZ black hole as well
as the Chern-Simons and dilatonic generalizations analyzed above are characterized by 
heat capacities which are positive in the physically meaningful interval. This means that 
these structures are thermodynamically stable and their corrections to the entropy can 
be analyzed by using to Eq.(\ref{corr}). 

\begin{figure}
%\begin{center}
\includegraphics[width=7cm]{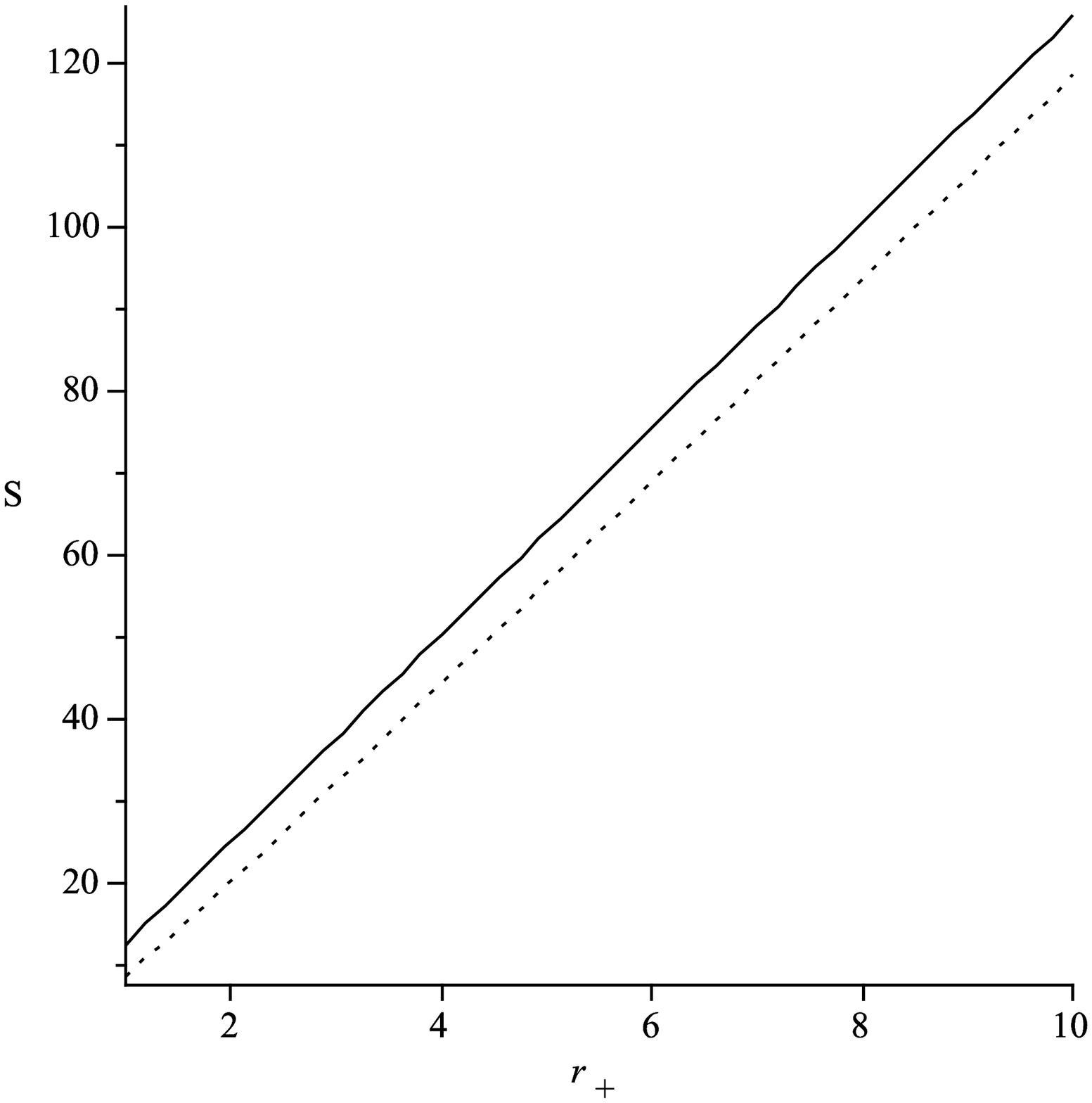}
\includegraphics[width=7cm]{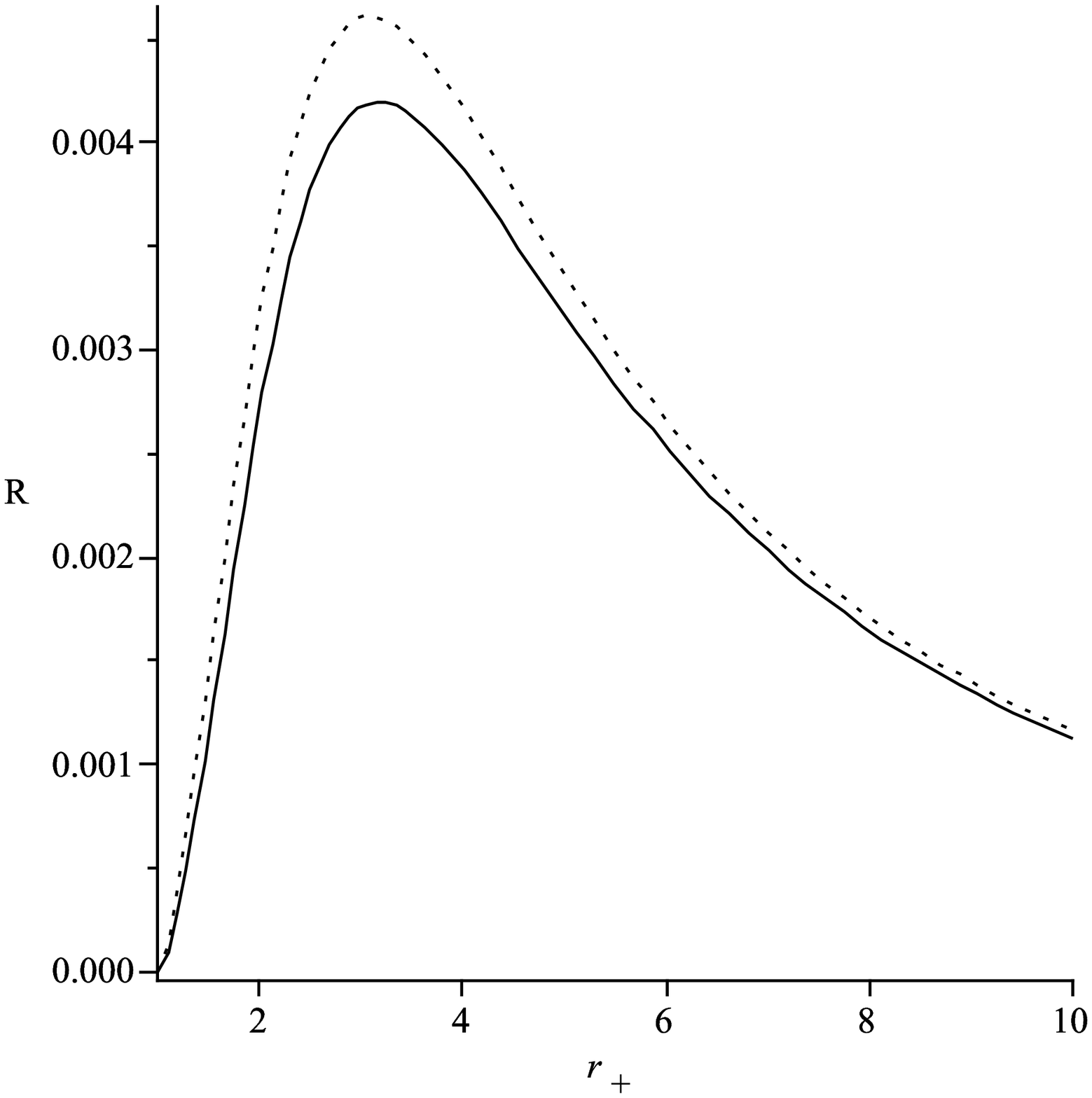}
%\end{center}
\caption{Behavior of the entropy and thermodynamic curvature for the BTZ black hole 
(solid curves) and the corresponding corrections (dashed curves). Our choice
of parameter values is $l=1$,  $r_-=1$ and $r_+> 1$. }
\label{fig:btzfluc}
\end{figure}

For the purposes of the present work, it is interesting to verify whether GTD is able to
correctly handle entropy corrections in the sense that a small perturbation of the entropy 
would correspond to a small perturbation of the thermodynamic curvature. To investigate this
question it is necessary to formulate GTD and the above results in the entropy representation. 
One of the advantages of GTD is indeed its flexibility in regard to different representations. 
We use as intuitive guidance the first law of thermodynamics in the entropy representation, i.e., 
$dS=(1/T)dM -(\Omega/T)d J$. Then, the fundamental equation should read $S=S(M,J)$ and 
the conditions for thermodynamic equilibrium are
$\partial S/ \partial M = 1/T$ and $\partial S/\partial J = -\Omega/T$ so that $1/T$ and $-\Omega/T$
are the variables dual to $M$ and $J$, respectively.  Consequently, the coordinates for 
the thermodynamic phase space ${\cal T}$ of the BTZ black hole can be chosen as 
$\{S,M,J,1/T,-\Omega/T\}$, and the fundamental form is $\Theta_S =dS - (1/T)dM +(\Omega/T)d J$. Applying 
the same prescription used to construct the metric (\ref{gup}), we obtain the following metric
for the phase space in the entropy representation
\be
G_S = \left(dS - \frac{1}{T} dM +\frac{\Omega}{T} d J\right)^2 + \left(\frac{M}{T} - J\frac{\Omega}{T}\right)
\left[-dM d\left(\frac{1}{T}\right) - dJ d\left(\frac{\Omega}{T}\right)\right] \ .
\ee
As for the coordinates of the space of equilibrium states ${\cal E}$, the natural choice is $\{M,J\}$ so that
the smooth map $\varphi: {\cal E}\longrightarrow {\cal T}$ with the condition $\varphi^*(\Theta_S)=0$ implies
the fundamental equation $S=S(M,J)$ and the equilibrium conditions given above. It is then easy to compute
the metric $g_S=\varphi^*(G_S)$ which can be written as
\be
\label{gdowns}
g_S = \left(M\frac{\partial S}{\partial M} + J\frac{\partial S}{\partial J}\right)
\left(-\frac{\partial^2 S}{\partial M^2} d M^2 
+ \frac{\partial^2 S}{\partial J^2} d J^2 \right)\ .
\ee
This metric represents the geometry of the equilibrium space for any thermodynamic system with fundamental equation $S=S(M,J)$. 
In the particular case of the BTZ black hole it reads
\be
\label{entbtz}
S_0= 2\sqrt{2}\ \pi l \left[ M +
\left(M^2 -\frac{J^2}{{ l}^2}\right)^{1/2}\right]^{1/2}\ .
\ee
Then, using the relationships (\ref{mj}), 
the corresponding thermodynamic metric can be expressed as
\be
g_S = \frac{2\pi^2l^4r_+^2(r_+^2+3r_-^2)}{(r_+^2-r_-^2)^3}
\left(dM^2 -\frac{1}{l^2}dJ^2\right)\ ,
\ee
from which it can be shown that 
the thermodynamic curvature in the entropy representation becomes
\be
\label{curvbtzs}
R_S = \frac{(r_+^2-r_-^2)^2 (5r_+^4 - 6r_+^2r_-^2 + 9r_-^4)}
{4\pi^2 r_+^4(r_+^2+3r_-^2)^3}\ .
\ee
We notice that this curvature essentially reproduces the results obtained in Section
\ref{sec:gtd} for the BTZ black hole geometrothermodynamics in the mass representation.
In figure \ref{fig:btzfluc} the general behavior of the curvature 
(\ref{curvbtzs}) is shown for a specific choice of the parameters. It can be seen that
the curvature is free of singularities in the entire interval, indicating that it 
corresponds to a stable thermodynamic system. 

In order to investigate in GTD the entropy correction for the BTZ black hole, 
we must introduce the entropy (\ref{entbtz}) into the general expression for the 
logarithmic correction (\ref{corr}), and the resulting corrected entropy must be
inserted into the general form of the thermodynamic metric (\ref{gdowns}) from which 
the  corresponding curvature can be derived in the standard manner. The resulting
expressions cannot be written in a compact form. Therefore, we perform a graphical 
analysis. The results are presented in figure \ref{fig:btzfluc}. The graphics
show clearly that a small correction of the entropy leads to a small perturbation of
the thermodynamic curvature. We performed similar analysis for the BTZ-CS 
and BTZ dilatonic black holes. The thermodynamic metrics of the corresponding 
equilibrium spaces in the entropy representation can be derived as described 
above for the BTZ black hole. In general, we conclude that a small perturbation 
at the level of the entropy corresponds to a small perturbation at the level
of the thermodynamic curvature. We interpret this result as a further indication
that in GTD the thermodynamic curvature can be used as a measure of thermodynamic 
interaction.

%%%%%%%%%%%%%%%%%%%%%%%%%%%%%%%%%%%%%%%%%%%%%%%%%%%%%%%%%
%%%%%%%%%%%%%%%%%%%%%%%%%%%%%%%%%%%%%%%%%%%%%%%%%%%%%%%%
\section{Conclusions}
\label{sec:con}

In this work we used the formalism of GTD to 
construct a thermodynamic metric for the space of equilibrium states 
of the BTZ black hole and its generalizations which include an
additional Chern-Simons term and a dilatonic field. In all theses 
cases we showed that the thermodynamic curvature is in general 
different from zero, indicating the presence of thermodynamic 
interaction, and free of singularities, indicating the absence of
phase transitions. This result is in accordance with the goals of 
GTD and allows us to investigate the thermodynamic properties of
the BTZ black holes in terms of the geometric properties of the 
corresponding space of equilibrium states. 

The thermodynamic metric proposed in this work has been applied to 
the case of black hole configurations in four and higher dimensions
with and without the cosmological constant \cite{aqs08,qs08}. In general,
it has
been shown that this thermodynamic metric correctly describes the 
thermodynamic behavior of the corresponding black hole configurations. One 
additional advantage of this thermodynamic metric is its invariance with 
respect to total Legendre transformations. This means that the results 
are independent of the thermodynamic potential used to generate the 
thermodynamic metric. Also, the generality of the method of GTD allows us 
to easily implement different thermodynamic representations. In particular, in the
case of BTZ black holes we presented in Section \ref{sec:gtd} the mass 
representation and in Section \ref{sec:fluct} the entropy representation.

For all BTZ black holes analyzed in this work, we showed that small perturbations 
at the level of the thermodynamic potential lead to small perturbations at the 
level of the thermodynamic curvature. This is not a trivial result that contrasts
with the results obtained by using other metric structures. In fact, Weinhold's metric
leads to a zero thermodynamic curvature for the BTZ black hole and to big 
perturbations of the curvature when small perturbations of the entropy are taken 
into account \cite{sarkar06}. 
This is not in agreement with the idea of describing thermodynamic
interaction in terms of curvature, which is one the aims of applying
geometric concepts in thermodynamics. In the case of GTD, the thermodynamic 
metric proposed for the equilibrium space not only leads no a 
non-zero thermodynamic curvature for the BTZ black hole, but also
induces small perturbations on the thermodynamic curvature when small perturbations
of the thermodynamic potential are taken into account. We interpret this result
as an additional indication that the thermodynamic curvature proposed in GTD can be 
used as a realistic measure for thermodynamic interaction.  

It would be interesting to further analyze the manifold of equilibrium states
of the BTZ black hole in the context of the variational principles proposed in GTD 
\cite{vqs08}. If it turns out that the thermodynamic metric investigated in 
the present work for the BTZ black holes satisfies the Nambu-Goto-like equations
\cite{pol}, an additional interpretation of BTZ configurations would emerge in terms of 
bosonic strings.  Recently, we analyzed the set of geodesics in the
equilibrium space of thermodynamic systems in the specific case of an ideal gas 
\cite{qsv08} and found a very rich geometric structure.  
It would be interesting to study the geometric structure of the space of 
geodesics in the case of the BTZ black which is one of the simplest black
hole configurations. 

In our geometric construction of GTD as presented above, 
the thermodynamic phase space plays only an auxiliary 
role in the sense that it is used only to correctly apply Legendre transformations 
and to guarantee Legendre invariance of metric structures. Nevertheless, it would be interesting
to analyze its geometric properties. We know, for instance, that its curvature
must be non-zero since a flat metric in the phase space is not Legendre invariant. 
It would be interesting to classify the phase space in terms of the properties of its curvature.
A preliminary 
study of its geodesic equations show that there exist solutions which contain information about 
the laws of black hole thermodynamics. This unexpected result must be deeply investigated 
in order to assure that the auxiliary phase space contains thermodynamic information.

\section*{Acknowledgements} 
This work was supported in part by Conacyt, Mexico, grant 48601.

% \appendix

% \section{}
% \label{}

%\appendix

%%%%%%%%%%%%%%%%%%%%%%%%%%%%%%%%%%%%%%%%%%%%%%%%%%%%%%%%%%%%%%%%%%%%% 
%%%%%%%%%%%%%%%%%%%%%%%%%%%%%%%%%%%%%%%%%%%%%%%%%%%%%%%%%%%%%%%%%%%%%%

\end{document}